\documentstyle[12pt,aasms4,epsfig]{article}

\begin{document}

\title {Atomic Carbon in Galaxies}

\author {Maryvonne Gerin,\altaffilmark{1,2} Thomas G.
Phillips,\altaffilmark{3}}

\altaffiltext{1}{Radioastronomie millim\'etrique -
 UMR 8540 du CNRS, Laboratoire de Physique de l'ENS, 24 Rue
Lhomond, 75231 Paris cedex 05, France.\\
maryvonne.gerin@physique.ens.fr}
\altaffiltext{2}{DEMIRM - UMR 8540 du CNRS, Observatoire de Paris,61 Avenue
de l'Observatoire 75014 Paris, France}
\altaffiltext{3}{Caltech Submillimeter Observatory, Caltech 320-47,
Pasadena, CA 91125.\\
phillips@submm.caltech.edu}

\abstract {We present new measurements of the ground state fine-structure line
of atomic carbon at 492 GHz in a variety of nearby external galaxies, ranging 
from spiral to irregular, interacting and merging types. In comparison
with  CO(1-0) emission
observed at the same spatial resolution, the CI(1-0) line intensity stays
fairly comparable in the different environments, with an average value of the
ratio of the line integrated areas in Kkms$^{-1}$ of 
CI(1-0)/CO(1-0) = $0.2 \pm 0.2$. However, some
variations can be found within galaxies, or between
galaxies. Relative to CO lines (J=2-1, 3-2, 4-3), CI(1-0) is weaker 
in galactic nuclei, but stronger in  disks, particularly outside 
star forming regions.  Also, in NGC~891, the CI(1-0)
emission follows the dust continuum emission at 1.3mm extremely well along the
full length of the major axis where molecular gas is more abundant than 
atomic gas. Atomic carbon therefore appears to be a good tracer of molecular
 gas in external galaxies, possibly more reliable than CO.

Atomic carbon can contribute significantly to the thermal budget of
interstellar gas. The cooling due to C and CO are of the same order of
magnitude for most galaxies. However, CO is generally a more important
coolant in starburst galaxies. Cooling due to C and CO  amounts
typically to $2 \times 10^{-5}$ of the FIR continuum or 5\% of the 
CII line. However, C and CO cooling  reaches 
$\sim$ 30\% of the gas total, in Ultra Luminous InfraRed Galaxies
(ULIRG) like Arp~220, where CII is abnormally faint.

Together with CII/FIR, the emissivity ratio CI(1-0)/FIR can be used as a
measure of the non-ionizing UV radiation field in galaxies. 
The plots of CII/CI or CII/FIR versus CI/FIR show  good correlations, 
in agreement with PDR models, 
except for two remarkable galaxies Arp~220 and Mrk~231, where high opacities 
of the CII line and possibly the dust thermal emission may be factors 
reducing the CII strength below the predictions of the
current PDR models.}

\section {Introduction} 
Due to the poor transparency of the Earth's
atmosphere at submillimeter wavelengths, there are few published measurements
of the ground state fine-structure lines of atomic carbon at 492 and 809 GHz
in external galaxies, despite the high abundance of this species in cool
interstellar gas and its importance for the overall thermal budget of
molecular gas. The first detection was reported by B{\"u}ttgenbach et al. 
(\cite{bkp}) in IC~342. A handful of galaxies have been detected
since, including NGC~253 (Israel et al.\cite {iwb}, Harrison et al. \cite 
{harrison}), M~82 (Schilke et al. \cite {schilke}, Stutzki et al. \cite 
{stutzki}), M~83 (Petitpas and Wilson \cite {pw}), M~33 (Wilson 
\cite {wilson}). Individual clouds have been observed in M~31
(Israel et al. \cite {itb}) and the LMC (Stark et al. \cite {starck}).

Atomic carbon can be found in all types of neutral clouds, from diffuse clouds
(Jenkins \& Shaya 1979) to dense molecular gas (Phillips
\& Huggins 1981). In diffuse clouds, atomic carbon is
a minor constituent, but the intensity ratio of the two ground
state fine-structure lines is a sensitive tracer of the total gas pressure.
Atomic carbon has also proven to be a good tracer of molecular gas in Galactic
molecular clouds, as a linear correlation is commonly found between the
strength of the  CI($^{3}P_1 - ^{3}P_0$) line at 492 GHz and the 
$^{13}$CO(2-1) line at 220 GHz (Keene et al. 1996). This
correlation corresponds to a mean abundance of neutral carbon of 
$\sim 10^{-5}$ relative to H$_{2}$. By comparing with extinction
measurements, Frerking et al. (\cite {frerking})
found a similar value for the abundance of atomic carbon in
dense molecular clouds, where it reaches a
maximum of $2.2 \times 10^{-5}$ for A$_{V}$ = 4 -- 11 mag and does not
deviate from this value by more than a factor of a few for larger A$_{V}$.
Emission from the two ground state fine-structure lines of atomic carbon is
seen by COBE throughout the Milky Way and makes a significant contribution to
the gas cooling (Bennett et al. 1994). Therefore, it is not surprising
 that these ubiquitous atomic carbon lines are present in the spectra 
of external galaxies.

From a theoretical point of view, the spatial distribution and line
intensities of atomic carbon in molecular clouds are predicted by chemical
models which include the effect of photodissociation induced by UV photons,
the so-called PDR models (e.g. Tielens and Hollenbach 1985). 
In clouds exposed to UV radiation, carbon is mostly
in the form of C$^+$ to a depth Av = 1 magnitude. Atomic carbon
appears at an intermediate depth, from Av = 1 to 5 or so magnitudes, where
C$^+$ is mostly recombined with electrons and not all the gas phase carbon
present is yet captured into CO. The actual depth of this zone, and thus the
extent and intensity of the CI emission, is a sensitive function of
some important model parameters, such as the carbon and oxygen abundance in
the gas phase, and the presence of Polycyclic Aromatic Hydrocarbons (see
Bakes and Tielens \cite{bt}, Le Bourlot et al. 1993). PDR models are difficult to
use for quantitative results in galaxies, because they were developed to
represent the structure of an individual cloud, whereas a galaxy would have to
be synthesized from a suitable ensemble of PDR clouds (see e.g. Sauty et al.
\cite{sauty}). Nevertheless, their predictions provide a qualitative
understanding of the variations of the line flux with physical
conditions.  Previous CII measurements of external galaxies have
been compared successfully with such PDR models (e.g. Stacey et al. 1991).

Although CII, CI and CO emission, in principle, arise from different physical
regions in the cloud, in actual observations of Galactic clouds it is found
that CI and CO emission on average seem to come from the same physical region.
This is in part due to the clumpy or fractal nature of the clouds, so that the UV
causes emission from irradiated regions even deep into the cloud. The overall
impression, at moderate (say 15'' at 500 pc = 0.04 pc) resolution, is that the
species are coexistent. This scale must be compared to the size of Galactic
clouds, which may vary from a few parsecs to a few tens of parsecs. 
Few large scale maps of
Galactic clouds have been made for all these species, but for some 
giant molecular clouds (Plume et al.  1994, 1999) there
is a good correspondence between the CI, CII and CO maps. 
For external galaxies where individual clouds are barely resolved,
we would then expect that the
different carbon species will be observationally coexistent (when 
associated with molecular gas) including the
ionized carbon. Therefore, at least for the point of view of beam
dilution in the observations, the 3 species (CII, CI and CO) can be considered
 as  occupying
the same volume in the beam so that observed line ratios can be compared with
model predictions. The ionized interstellar medium, either diffuse or in
HII regions, can contribute to the global CII emission of galaxies, 
but for starforming regions in galaxies, PDRs contribute
most of the CII emission (Madden et al. 1993, Sauty et al. 1998). For
M~82 the contribution of HII regions is estimated to be $\sim 24\% -
31\%$ of the observed CII flux (Colbert et al. 1999).
 
The contribution of the cool diffuse regions to the atomic carbon
emission will be incorporated in
the PDR models. The warm diffuse gas is generally ionized and
therefore barely contributes to the CI emission.
Therefore, the two fine-structure lines of atomic carbon 
are expected to be better tracers of PDRs than is CII.
The intensity ratio  CI(1-0)/CO(1-0) is a function of the gas density
and UV illumination factor G$_{0}$, as shown in Figure 1, produced
with the PDR model developed by Le Bourlot et al. (1993) and Abgrall
et al. (\cite{aa}). 
  G$_0$ is defined relative to the average interstellar radiation field in 
the solar neighborhood as obtained by Mathis et al. (\cite{mmp}) G$_{0}$ = 
I$_{UV}$(6-13.6 eV)/ $1.4 \times 10^{-4}$ erg\ cm$^{-2}$s$^{-1}$sr$^{-1}$;
 n$_H$ represents the total density of hydrogen atoms.

Other ratios, CII/CI,  CI/FIR and CII/FIR depend mostly
on the UV illumination factor G$_0$ (see also Kaufman et al. 1999).
Since atomic carbon is present in a layer of the PDR
with almost the same characteristics (column density, temperature)
whatever the illumination, the brightness of the CI(1-0) line is fairly 
constant for most of the parameter space studied. This means that the
main factor affecting this line in external galaxies is the filling factor 
of the emission in the beam. By contrast,  
the FIR emission in a PDR is directly proportional to the UV illumination since
nearly all the incident UV radiation is absorbed in the cloud. Finally, the CII and
OI emission in PDRs, depend on both the UV illumination (due to
the decrease of the efficiency of photoelectric heating at
high illumination) and the gas density (for the collisional excitation
of these lines). From the PDR models, it is expected that the CI/FIR ratio 
will be a useful tracer of the local UV illumination conditions.

In the following we present new observations of CI(1-0), as well as
complementary data, $^{13}$CO(2-1), $^{12}$CO(2-1), (3-2) and (4-3), 
on nearby galaxies, of various
morphological types. These data are used to study the correlation of CI
emission with other tracers of the ISM in galaxies.

\section{Observations} 
The observations were performed with the Caltech
Submillimeter Observatory (CSO) during 1996 - 1998, using SIS receivers
operated in double-side (DSB) mode. In good atmospheric conditions 
($\tau_{225 GHz} \leq 0.06$), the single sideband system temperature usually 
ranged between
1000 and 3000 K. The observations were performed using a chopping secondary
mirror, with throw set to 1 to 3 arcminutes on the sky, depending on the size
of the source. The spectra were analysed with two acousto-optic spectrometers,
one with a total bandwidth of 1500 MHz (of which only 900 MHz is used, due to
the bandwidth limit of the receivers) and a resolution of about 2 MHz, and a
second with a total bandwidth of 500 MHz and a spectral resolution of about 1.5
MHz. The main beam efficiencies of the CSO were 0.72, 0.65 and 0.53 at 230,
345 and 492 GHz respectively; the corresponding conversion factors between
Janskys and Kelvins (T$_A^*$) are 50, 70 and 100 Jy/K, 
and beam sizes 30 \arcsec, 20\arcsec and 15 \arcsec. Some data
have been reported previously (Gerin \& Phillips \cite {gp2,gp1}). 
The program galaxies are listed in Table 1.

Most of the target galaxies have narrow lines which fit in a single backend
spectrometer setting. For two galaxies with broad lines, Arp~220
and the center of NGC~3079, we used three different local
oscillator settings to obtain a complete coverage of the lines, with central
velocities at -300, 0 and +300 kms$^{-1}$ relative to the line
center. First, we used the spectra taken at velocity offsets +300 and -300
kms$^{-1}$ to adjust the zero level of the baseline, with a
window set by reference to published CO spectra. We then computed the baseline
offset for the central spectrum, to minimize the platforming effects with the
spectra taken at +300 and -300 kms$^{-1}$. The overlap region is
quite large, +70 to +200 kms$^{-1}$ at positive velocities for
Arp~220 for example. The data and interpretation for
Arp~220 have been reported elsewhere (Gerin \& Phillips \cite {gp1}).

\section {Results}
CI(1-0) was detected in all the program galaxies. Figure 2 presents a
selection of CSO spectra. The  measured parameters are listed in Table
2, together with
complementary data found in the literature.   Linear baselines have
been fitted with the original spectral resolution ($\sim 1.5$ MHz =
0.8 kms$^{-1}$). Depending on the expected line width, the spectra
have then been rebinned to 5  -- 15 kms$^{-1}$ resolution to increase 
the S/N ratio. For galaxies with broad lines
and not too many channels beyond the line, we 
chose the line window according to published CO spectra. The
uncertainty in the line flux due to the uncertainty in the
position of the baseline is included in the total uncertainty listed
in Table 2.  Figure 3 shows  the
integrated intensity ratio CI(1-0)/CO(1-0) in T$_{mb}$ as a function of
the CO(1-0) integrated intensity in Kkms$^{-1}$. This ratio  has
a mean value of $0.2 \pm  0.2$, with a large dispersion, since the observed
 ratios range from $\sim$ 0.04 to $\sim$1. To get the emissivity ratio, 
the observed intensity ratio in
T$_{mb}$ must be multiplied by 78, the cube of the ratio of the line 
frequencies. The mean emissivity ratio is therefore 16, with observed
values ranging from 3 to 80. In this figure the CO(1-0) data are taken from
the literature, as listed in Table 2. We also included 
any already published CI data. We used mostly the CO(1-0) data taken with the 
IRAM 30m telescope in
order to match the CI beam size at the CSO as closely as possible. 
The CO(1-0) data taken with
other telescopes have been scaled to the 22'' beam of the IRAM 30m telescope
at 2.6 mm. 

Despite the large scatter in the CI/CO ratio, there is no apparent segregation
between the different galaxy types. Viewed at large scale (the 15" CSO beam
represents a linear scale of 730 pc at a distance of 10 Mpc), there are no
clear differences between normal spirals, merger galaxies and even low
metallicity galaxies, except maybe two regions in M~33. However,
there are local differences within galaxies, which we examine in the following
subsection. Because $^{13}$CO emission is 
low in active galaxies (e.g. Taniguchi and Ohyama 1998),
 we compare  CI(1-0) with $^{13}$CO(2-1) in section 3.2. 
We then discuss  the respective contributions of atomic
carbon and carbon monoxide to the gas cooling in galaxies, and finally 
we compare the CI(1-0) emission of nearby galaxies with the CII and
 FIR emission at the same angular resolution.

\subsection {Disk galaxies} 
We have mapped the north-east half of the major axis of the edge-on
galaxy NGC~891, made a cut through the disk of NGC~6946, and observed points
in the nucleus and in the disk of several other galaxies. We  investigate the
difference between  galaxy nuclei and the disks, and between
active regions in spiral arms and the general interstellar medium.

As a point of
reference, we examine in Figure 4 the
distribution of CI and CO lines in the Milky Way Galaxy, as observed by FIRAS
on board COBE (Bennett et al. \cite{bennett}). It is clear that for the 
Milky Way, CI(1-0) decreases less rapidly than the CO lines with
Galactic radius, moving out along
the Galactic plane. The decrease is steeper for CO(4-3) than for CO(2-1).
Also, neutral carbon is more excited in the nucleus, as seen by the higher
CI(2-1)/(1-0) emissivity ratio : 1.3 in the nucleus versus 0.5 in the disk.
The 7$^\circ$ beam of COBE corresponds to a linear size 
of 1 kpc at the Galactic Center,
quite similar to the linear size of the 15'' CSO beam at a distance of 10 Mpc
(0.7 kpc). Also shown in Figure 4 is the dust continuum at 1mm from
the same COBE/FIRAS data. It is clear that the continuum intensity 
does not drop as fast as the CO and CI lines outside the nucleus of
the Milky Way.

Figure 5 and 6 present a comparison of the CI(1-0) intensity with CO(1-0) in
NGC 891 and NGC 6946 respectively.
A similar effect to that seen in the Milky Way is seen for 
NGC~891, NGC~6946 and also for NGC~3079 : CI(1-0) is
weaker, relative to CO(1-0) and (2-1), in the nucleus than in the disk. Toward
NGC~891, we can compare CI(1-0) with both CO(1-0) and the dust continuum
emission at 1.3mm from Gu\'{e}lin et al. (1993). Whereas the CI/dust continuum 
ratio stays constant
throughout the disk, the CO/dust continuum increases, and the CI/CO decreases
in the nucleus. Nuclear gas is usually denser and warmer than the bulk of the
interstellar gas in galactic disks, resulting in a higher excitation
temperature for carbon monoxide in the nucleus. This difference in excitation
has some consequences for the use of $^{12}$CO(1-0) as a tracer
of molecular gas since the mass of molecular gas deduced from $^{12}$CO(1-0)
 data will be overestimated in galaxy nuclei. This has been
shown for the Galactic Center by Sodroski et al. (1995).

For NGC~6946, the CI(1-0)/CO(2-1) ratio increases just outside the nucleus, in
an interarm region, and decreases again when reaching a star forming region
near the position (0,+140''). For other galaxies also : M~33  (Wilson 1997),
 IC~10 (this work), a similar difference is found between
clouds close to HII regions and clouds more distant from star forming
complexes.
CI(1-0) appears to be a tracer of the general interstellar medium,
not strongly biased toward the densest and warmest places.

\subsection {Comparison with $^{13}$CO(2-1)}
In the Milky Way, a remarkable linear correlation between CI(1-0) and
$^{13}$CO(2-1) intensities has been found in most of the clouds
mapped (Keene et al. \cite {keene}). This correlation can be understood in
the sense that both atomic carbon and $^{13}$CO are good tracers of the
 molecular gas given the
permeability of the clouds to the UV radiation : the two lines have similar
(moderate) opacities and are close to LTE for most of the physical conditions
encountered in local molecular clouds. However, compared
to maps of individual molecular clouds in the Milky Way,  larger linear scales
are sampled in external galaxies where the relation between CI and $^{13}$CO
has not yet been studied. It is well known that $^{13}$CO lines are especially 
weak relative to $^{12}$CO in some interacting and merging galaxies
(Casoli et al. \cite {casoli2}, Aalto et al.
\cite {aalto}, Taniguchi and Ohyama 1998). We have therefore looked at
possible variations of the CI/$^{13}$CO(2-1) ratio as a function
of the $^{13}$CO(2-1) intensity. The data are displayed in
Figure 7, with the same symbols as in Figure 3. We used $^{13}$CO(2-1)
 data from the literature as listed previously, or took new
CSO data when needed.

This plot shows a trend of decreasing CI/$^{13}$CO(2-1) ratio
with the $^{13}$CO intensity, with a few exceptions :
M~33, the nucleus of Centaurus A and NGC~253. The trend
is particularly clear for the disk of NGC~891 (black triangles in
Fig. 7). Such a trend has been observed towards well known PDRs (Tauber et
al. \cite {tauber}, White \& Sandell \cite {ws}, Minchin \& White
\cite {mw}). It can be understood as revealing a smooth variation of the
abundance ratio N(C)/N(CO) with the total gas column density. Indeed, chemical
models predict a decrease of the N(C)/N(CO) column density ratio with
increasing H$_2$ column density in the range $ 0.1 - 5 \times 10^{21}$ 
cm$^{-2}$, valid for translucent
clouds (Stark et al. \cite {stark-r}). If translucent gas represents a fair
 fraction of the
molecular gas in external galaxies, atomic carbon lines will trace
preferentially these physical conditions. As mentioned in the Introduction,
from our knowledge of interstellar clouds in the Milky Way, it is expected
that a significant fraction of the molecular gas is exposed to UV radiation.
 This seems to be the case in external
galaxies as well since the CI(1-0) line at 492 GHz shows up at a detectable
level in all types of galaxies, and does not show any peculiar behavior with
the morphological type. Furthermore, in NGC~891, CI emission closely follows 
the dust continuum emission, which is also believed to be a reliable gas
tracer in spiral galaxies. We conclude that CI(1-0) can be used as a tracer of
low to moderate density molecular gas in external galaxies. It is possible that 
CI avoids the regions with high  density, which represent a small fraction 
of the total gas mass of galaxies.

\section {Gas cooling in galaxies}

\subsection {C and CO cooling}
We have estimated the total cooling due to the observed lines of atomic
carbon, CI(1-0), and carbon monoxide, CO(1-0) - (2-1) - (3-2) and in some cases
(4-3). The data are reported in Table 3. It turns out that 
a significant fraction of the cooling is due to the CO(3-2) and (4-3) lines,
 so we report only a lower limit for CO
cooling in galaxies with no CO(3-2) or (4-3) data. 
In order to estimate the total cooling due to all
carbon and CO lines, we need to correct for the missing
lines : CI(2-1) at 809 GHz, and CO(5-4), (6-5),... As templates, we used
well studied cases, either the COBE data towards the Galactic Center or 
data on M~82 and IC~342 (G\"usten et al. \cite {gusten}). The correction 
for missing lines amounts, on average, to approximatively
a factor 2 for carbon and a factor 4 for CO when no
CO(4-3) and higher J data are available, or 2 when no CO(6-5) and higher J are
available.

Israel et al. (1995) conclude that 
the contributions of CO and atomic carbon to the total cooling are
similar for NGC~253. We show that this conclusion holds for all  
the  galaxies in this sample. The contribution of CO and
atomic carbon to the total cooling  represent a small fraction of the total gas
cooling, which is still dominated by CII, and possibly OI. 
Typically, the gas cooling due to atomic carbon or CO amounts to 
10$^{-5}$ of the FIR dust continuum, while the gas cooling due to CII
and OI  amounts to
10$^{-4}$ -- 10$^{-2}$ of the FIR dust continuum.

\subsection {Comparison with CII and FIR}

We show in Figure 8 the line to continuum ratio CII/FIR versus CI(1-0)/FIR,
and the line ratio CII/CI(1-0) versus CI(1-0)/FIR. 
The lines drawn in Figure 8  show PDR model predictions 
(using the code described in Le Bourlot et
al. 1993)  for hydrogen densities from 10$^{2}$ cm$^{-3}$ to
 10$^{5}$ cm$^{-3}$ and G$_{0}$ from 1 to 10$^{4}$ times the average radiation
field in the solar neighborhood. 
The Far-infrared emission (FIR) is calculated from the measured fluxes
at 60 and 100 $\mu m$, available from observations with the
Kuiper Airborne Observatory (Madden et al. 1997, 1998, Stacey et al. 1991,
Smith and Harvey 1996) or with the Infrared Space Observatory (ISO)
(Luhman et al. 1998) with the formula :
$FIR = 1.26 \times 10^{-14} (2.58 F_{60} + F_{100}) Wm^{-2}$, where
$F_{60}$ and $F_{100}$ are the fluxes at 60 and 100 $\mu m$ in Janskys.
Since the CII and FIR data have been taken at
a lower spatial resolution than the CI data, we have smoothed or extrapolated
 the CI(1-0) data to a 55'' beam,
the spatial resolution of the CII and FIR data.
For nearby spiral
galaxies (NGC~891, NGC~6946), the data correspond to the central 55''. The
CII/CI(1-0) ratio shows a remarkable decreasing trend with 
increasing  CI(1-0)/FIR ratio, except
for the two ultraluminous infrared galaxies, Arp~220 and Mrk~231. The line 
to continuum ratio
CII/FIR is also well correlated with CI(1-0)/FIR (Fig. 8) for all the 
studied galaxies, with the exceptions of
 Arp~220 and Mrk~231 which again lie significantly
lower than the PDR model predictions.

These two plots can be understood as showing variations of the mean UV
radiation field in the galaxy sample : for moderate UV illumination
(G$_{0}$ = 10 - 100), the gas heating due to the photoelectric effect on
grains reaches maximum efficiency (Kaufman et al. 1999), so that the line to
continuum ratios CI(1-0)/FIR and CII/FIR  reach their maximum values. 
When the
UV radiation field increases, the gas heating is less efficient, while the dust
heating stays at the same level per UV photon. In that case, the far 
infrared radiation scales linearly with G$_{0}$,
while the CII emission scales only logarithmically, and the CI(1-0) emission 
stays nearly at the same level. The ratio CI(1-0)/FIR and CII/FIR are
thus  expected to
decrease as G$_{0}$ increases, and can be used to get estimates of the
intensity of the radiation field in starburst galaxies. Because CI(1-0) is
less sensitive than CII to G$_{0}$ and the gas density $n_{H}$, a
variation of the line ratio CII/CI(1-0) is predicted by PDR models
(cf Figure 1), which explains the trends seen in Figure 8.

Most of the studied galaxies lie in the parameter space, G$_{0}$ =
10$^{2}$ to 10$^{3}$, n = 10$^{2}$ to 10$^{5}$, as expected for 
molecular clouds in central regions of nearby galaxies. In
M~82, a nearby starburst galaxy, the radiation field is more intense and
reaches $\sim 10^{3}$ in good agreement with the estimate given by Kaufman
et al. (1999). The two merger galaxies, Arp~220 and Mrk~231 lie at the bottom 
left of the CII/FIR versus CI(1-0)/FIR plot,
where both the radiation field and the gas density are high. For these
extremely bright sources, the emission of individual PDR clouds can not be
simply added as in nearby spiral galaxies and the opacity of the CII and 
OI lines (and possibly also the opacity of the dust thermal emission) must
be taken into account as discussed in Gerin and Phillips (1998) and Luhman
et al. (1998).

Because most of the PDR CII and CI(1-0) emission arises from the surface of molecular
clouds, the filling factors of the CII, CI and FIR emission are expected to be
similar for starburst galaxies. Furthermore, since CI(1-0) is less likely than CII
to be contaminated by other emission sources (HII regions, WIM etc.), the
CI(1-0)/FIR ratio is a better measure of G$_{0}$ than the CII/FIR ratio. For
star forming galaxies, where the dust heating is dominated by the radiation of
massive stars, the line to continuum ratio CI(1-0)/FIR can thus be used to measure
the strength of the UV radiation field. 
This conclusion is valid for spiral galaxies with the same
metallicity as the Milky Way. In low metallicity galaxies, the 
line and continuum emission do not scale the same way with
metallicity and specific models must be used (Lequeux et al. 1994, 
Pak et al. 1998).

\section {Conclusions}

We have shown that the ground state fine-structure line of neutral carbon can
be detected in a variety of  external galaxies with  current instrumentation.
CI(1-0) data can be used in addition to other measurements to give information
on the gas column density, the thermal balance and the UV illumination
conditions in external galaxies. The results will be useful as 
templates for distant galaxies, when detectable with advanced instrumentation.

In nearby galaxies, atomic carbon is weaker relative to CO in the nucleus,
but more widely distributed in the disk.  This has important
consequences for the thermal balance of the molecular interstellar medium.
As a whole, the contribution of C and CO to the gas cooling
are of the same order of magnitude :  $2 \times 10^{-5}$ of the FIR continuum 
or 5\% of the 
CII line. C and CO cooling becomes significant, reaching 
$\sim$ 30\% of the gas total, in merger galaxies like Arp 220
where CII is abnormally faint. 
This conclusion rests on the current data where only the ground state
line of atomic carbon has been observed. Data on the second fine-structure
line of atomic carbon at 809 GHz and on other CO lines are needed to 
obtain a better measurement of the gas cooling in 
galactic disks and nuclei.

\acknowledgments { We are grateful to F. Boulanger and G. Lagache
for helping us using the COBE/FIRAS data. We thank J. Le Bourlot, 
G. Pineau des For\^ets and
E. Roueff for the use of their PDR model, and M. Gu\'elin
for sending us the continuum map of NGC~891. The CSO is
funded by NSF contract AST96-15025. M. Gerin acknowledges travel grants from
INSU/CNRS, and NATO. }




\begin{center}
{\large FIGURE CAPTIONS}
\end{center}

\begin{figure} [ht]
\epsfig{file=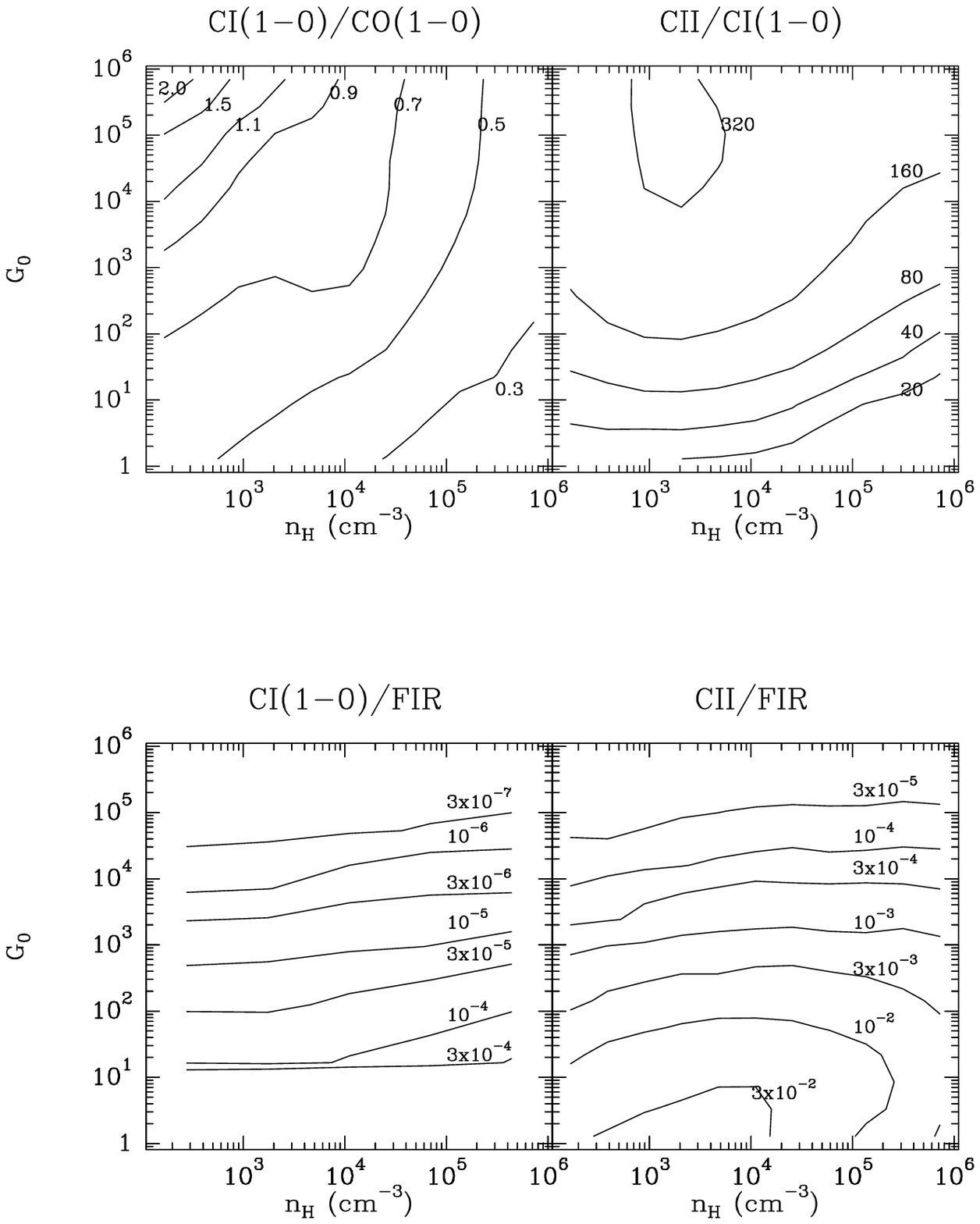,height=20cm,width=16cm}
\caption {
(Top left): Comparison of the intensity ratio CI(1-0)/CO(1-0) 
(in Kkms$^{-1}$)  for a face-on PDR as functions of
the uniform gas density n$_H$ and UV illumination factor G$_0$. 
Contours are drawn at 0.3, 0.5, 0.7, ... 2. (Top Right) : Same comparison for
CII/CI(1-0) (in erg\ cm$^{-2}$s$^{-1}$sr$^{-1}$). 
Contours are drawn at 20, 40, 80, 160, ... 640.
(Bottom left) :  Same comparison for
CI(1-0)/FIR (in erg\ cm$^{-2}$s$^{-1}$sr$^{-1}$). 
Contours are drawn at $3 \times 10^{-7}$, $10^{-6}$,
$3 \times 10^{-6}$, ... $3 \times 10^{-4}$.
(Bottom right) :  Same comparison for
CII/FIR (in erg\ cm$^{-2}$s$^{-1}$sr$^{-1}$). Contours are drawn at 
$3 \times 10^{-5}$, $10^{-4}$, $3 \times 10^{-4}$, ... $3 \times 10^{-2}$.}
\end{figure}

\begin{figure}[ht]
\vspace {20cm}
\includegraphics{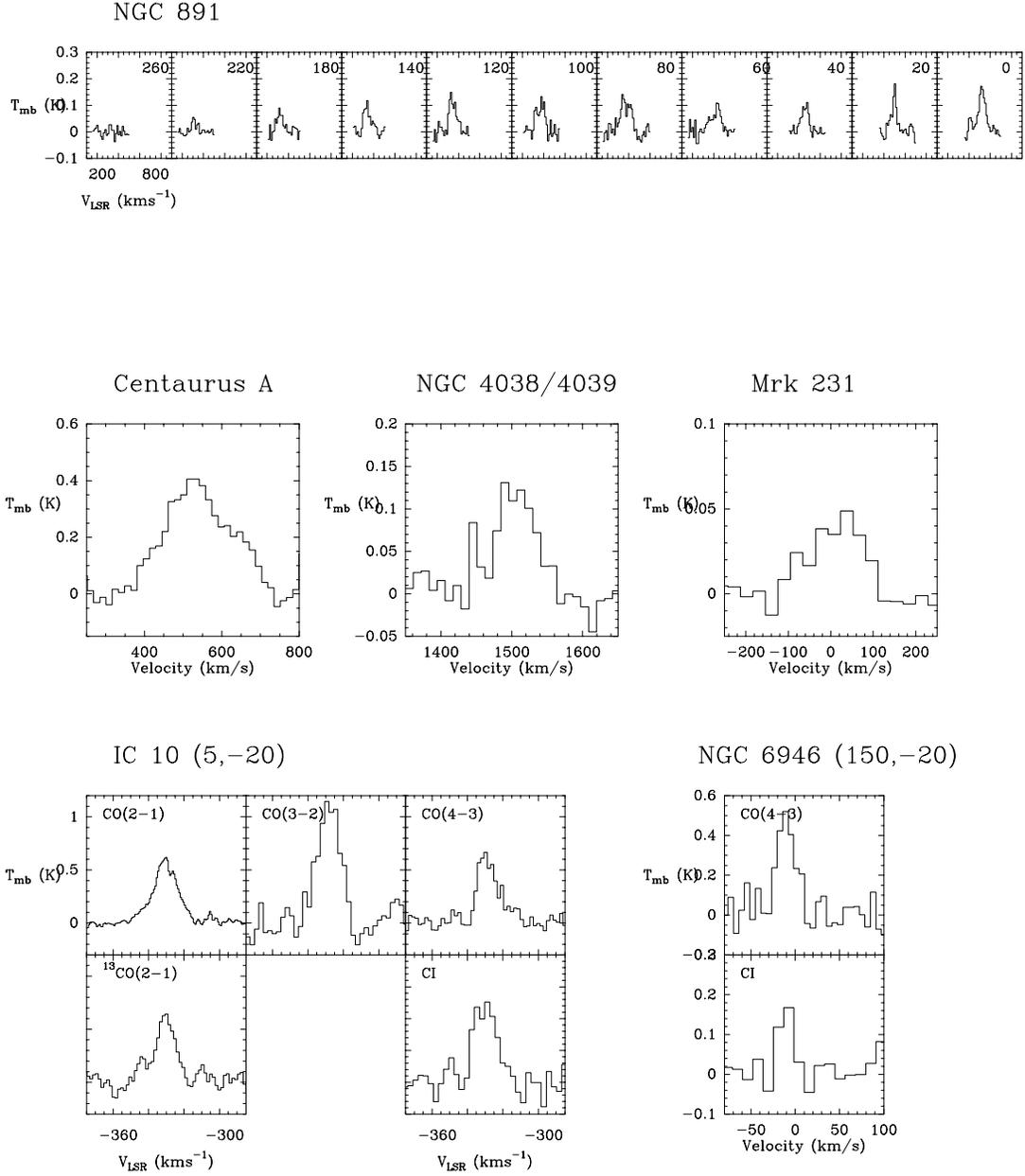}
\caption{A selection of CSO spectra. (Top) : CI(1-0) data taken
along the disk of NGC 891, (Middle) : Centaurus A (central region),
The Antennae (overlap region) and Markarian 231,
, (Bottom) : CO(2-1), (3-2), (4-3),  $^{13}$CO(2-1) and
CI(1-0) towards the nucleus of IC 10 (offsets (5'',-20'')
from the position in Table 1), CO(4-3) and CI in a
star forming complex in the disk of NGC~6946 (offsets (150'',-20'')
from the position in Table 1). The temperature unit is T$_{mb}$ (K),
velocities are given relative to the LSR.}

\end{figure}
      
\begin{figure}[ht]
\epsfig{file=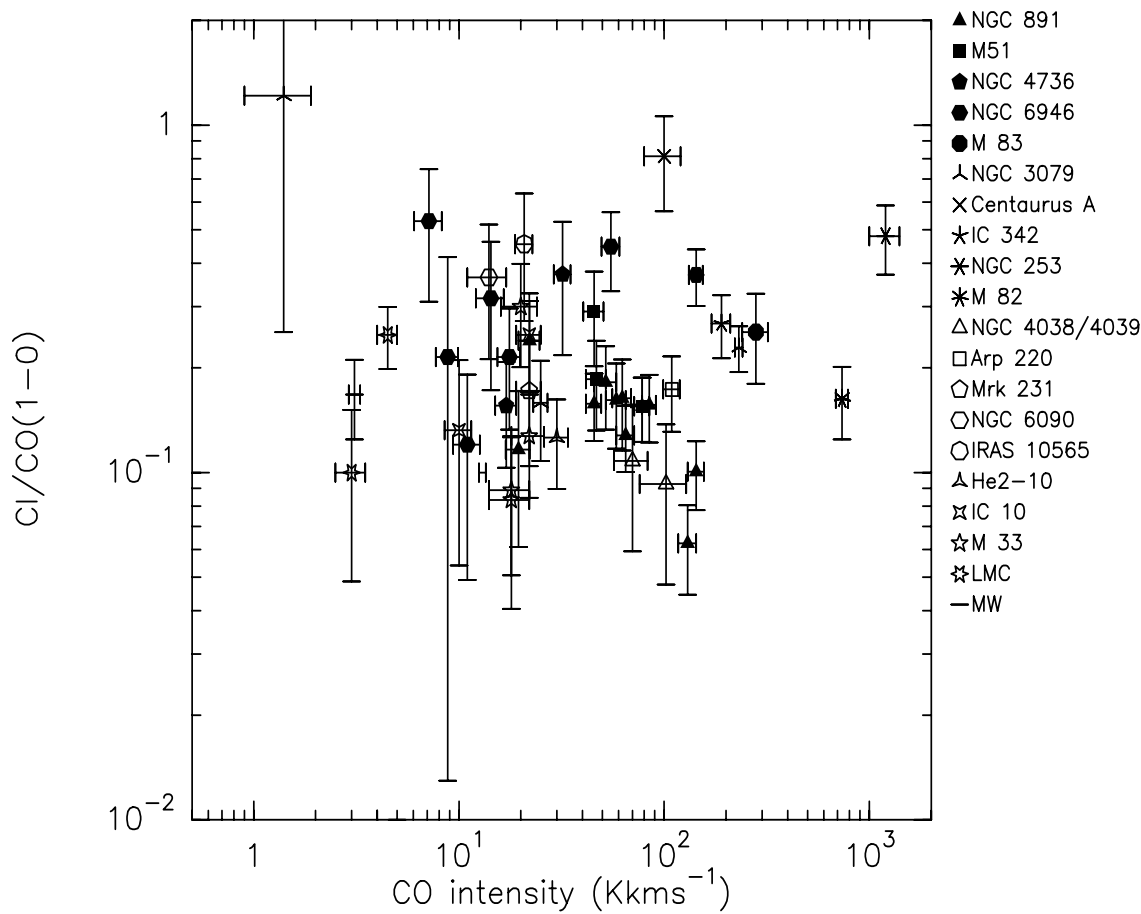,height=12cm,width=15cm,angle=0}
\caption {CI(1-0)/CO(1-0) Intensity ratio in Kkms$^{-1}$ as a
function of the CO(1-0) intensity in a 22" beam. The line intensities are
given in T$_{mb}$. Different markers are used according to the galaxy
type : black markers for normal spirals, cross-like for starburst and
active galaxies, white for merger and interacting galaxies, and star-like
markers for low metallicity galaxies. There is no obvious grouping by galaxy type in this presentation.}
\end{figure}

\begin{figure} [ht]
\epsfig{file=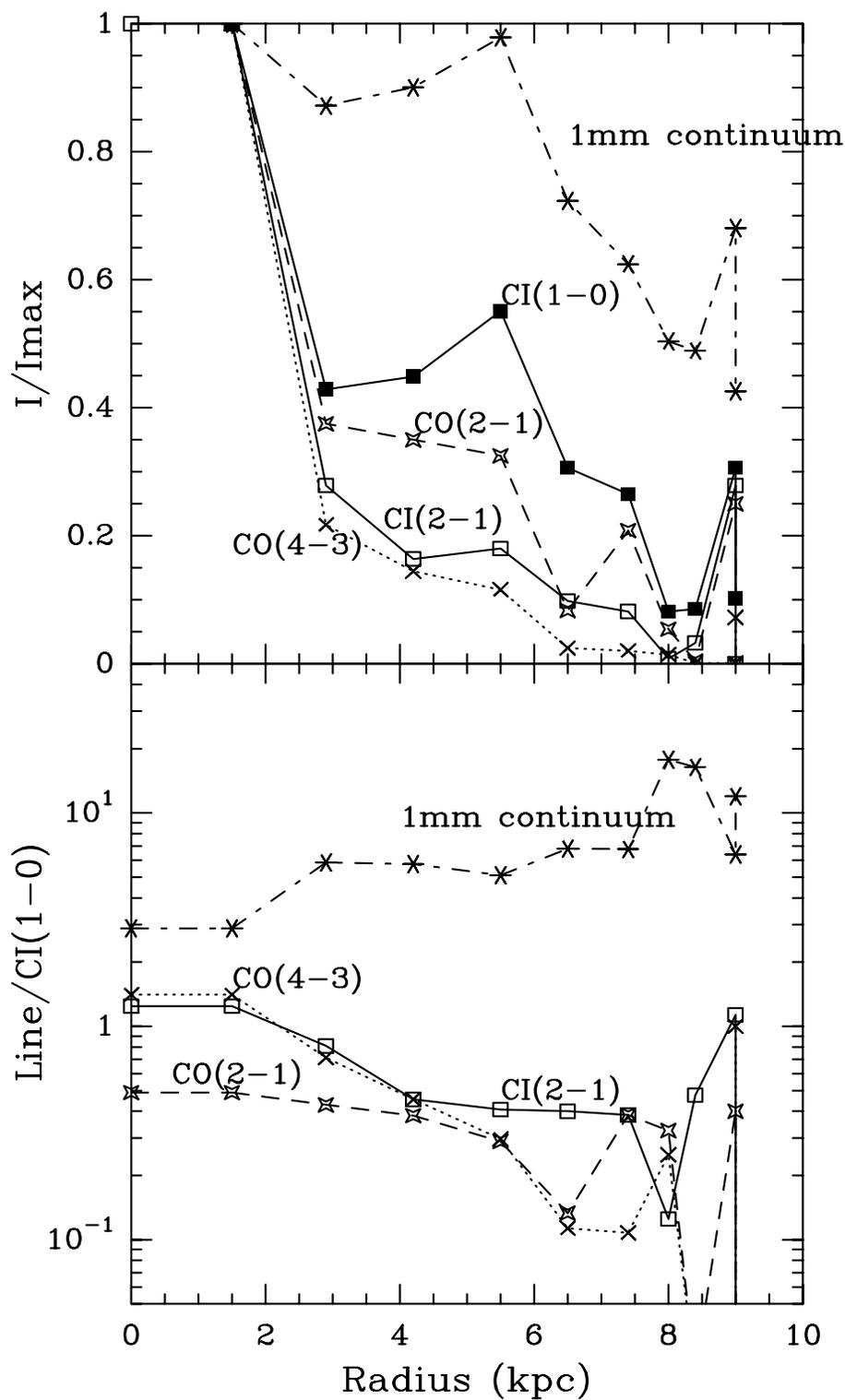,height=20cm,width=12cm,angle=0}
\caption {Upper : Distribution of the CI(1-0) (bold line, black squares),
CI(2-1) (thin line, white squares), CO(2-1) (dashed line, stars) and CO(4-3)
(dotted line, crosses) along the Galactic plane from COBE-FIRAS data.
The 1mm continuum intensity is also shown for comparison (dotted-dashed
line with stars). The intensity
is normalized to the peak intensity at the Galactic Center. Lower : Emissivity
ratio of CI(2-1)/CI(1-0) (full line, white squares), CO(2-1)/CI(1-0) (dashed
line, stars) and CO(4-3)/CI(1-0) (dotted line, crosses) along the Galactic 
plane from COBE-FIRAS data.  The intensity ratio 1mm continuum/CI(1-0)
 is also shown (dotted-dashed line 
with stars). We have converted the Galactic longitude to distance
from the Galactic Center assuming that the emission occurs at the tangent
point, and using a distance of 8.5 kpc to the Galactic Center.}
\end{figure}

\begin{figure}[ht]
\epsfig{file=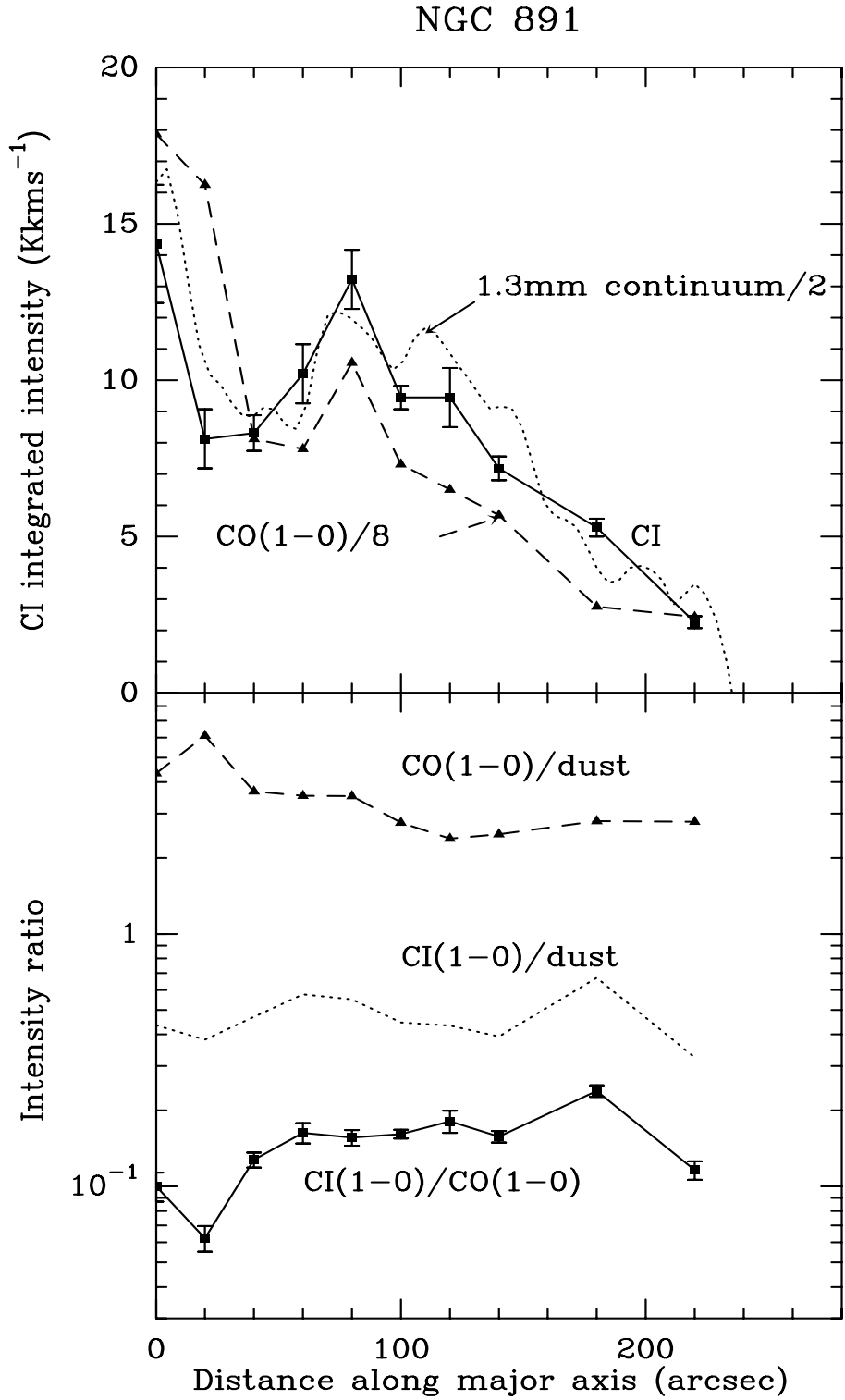,height=20cm,width=12cm}
\caption {Upper : Distribution of the CI(1-0) (full line), CO(1-0) (dashed
line, Garcia-Burillo et al.
\cite {gg}) and dust continuum (dotted line, Gu\'elin et al.
\cite {guelin}) along the major axis of NGC~891. The units of CI and
CO(1-0) are Kkms$^{-1}$ and the dust continuum is in mJy. Lower : Intensity
ratios along the major axis of NGC~891.}
\end{figure}

\begin{figure}[ht]
\epsfig{file=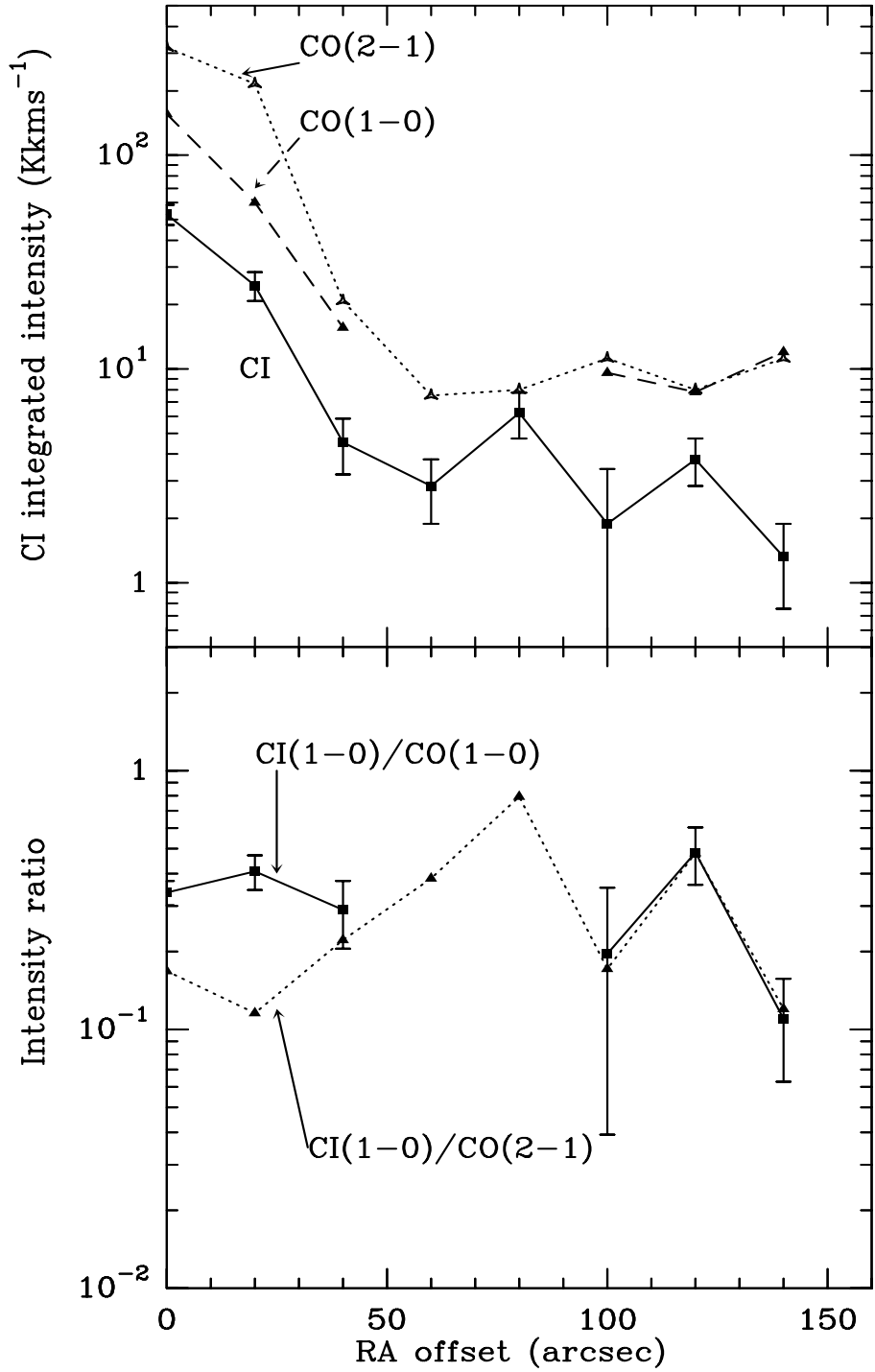,height=20cm,width=12cm}
\caption {Upper : Distribution of the CI(1-0) (full line), CO(1-0) (dashed
line, Casoli et al. \cite {casoli}) and CO(2-1) (dotted line, Sauty et al.
\cite {sauty}) in the disk of NGC~6946. Lower : Intensity ratio
for these lines. } 
\end{figure}

\begin{figure} [ht]
\epsfig{file= 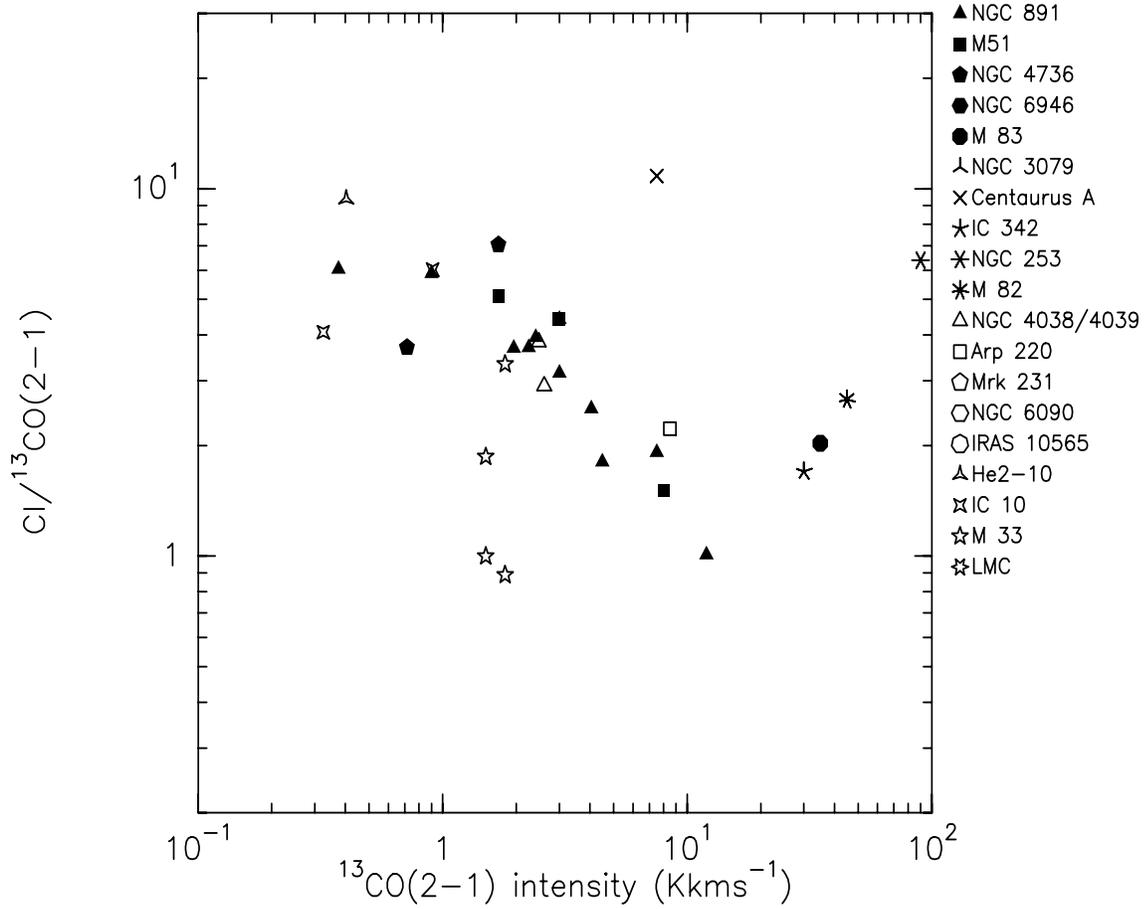,height=12cm,width=15cm,angle=0}
\caption { CI/$^{13}$CO(2-1) intensity ratio as a function of
the $^{13}$CO(2-1) intensity. Symbols are the same as for
Fig. 3.}
\end{figure}

\begin{figure} [ht]
\vspace {20cm}
\includegraphics{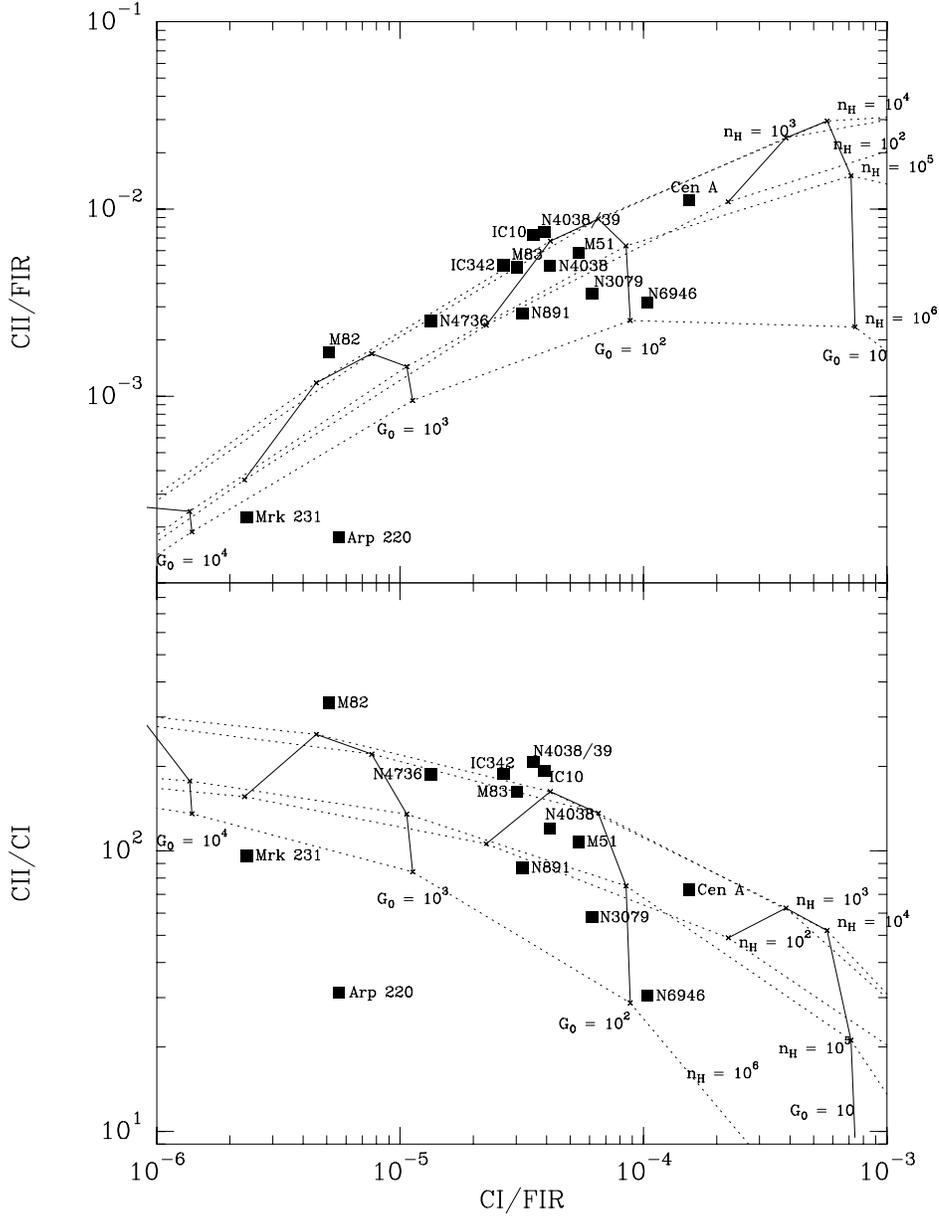}
\caption {Line to continuum ratio CII/FIR (upper) and line
emissivity ratio CII/CI (lower) versus CI/FIR in a 55" beam. The
lines show PDR model predictions for n$_H$ = 10$^{2}$ to
10$^5$ cm$^{-3}$, and G$_{0}$ = 10 to
10$^{4}$. The full lines are for constant G$_0$; the dashed lines for
constant density.} 
\end{figure}


\newpage
\vfill
\eject

\begin{table}
\caption{Parameters of the observed sources}
\begin{tabular}{lcrrrrl} 
\hline
 Name & Type & RA(1950) & Dec (1950) & V$_{LSR}$ ($cz$) & D(Mpc) &
Comments \\
    &        &         &            & kms$^{-1}$ & & \\ 
\hline 
NGC 891 & Sb & 02:19:24.7 & 42:07:18.6 & 535 & 9.5 & 11 positions \\
NGC 3079 & S & 09:58:35.0 & 55:55:15.4 & 1331 & 16 & 3 positions \\
NGC 4736 & Sab & 12:48:31.9 & 41:23:31.2 & 314 & 5 & 2 positions \\
M 51 & Sb & 13:27:46.1 & 47:27:14 & 470 & 9.6 & 4 positions \\
NGC 6946 & Sc & 20:33:48.8 & 59:58:50.0 & 50 & 5 & 10 positions \\
Centaurus A & E/Irr & 13:22:31.65 & -42:45:32 & 550 & 3 & Nucleus \\
NGC 4038 & Int. & 11:59:19.0 & -18:35:23 & 1634 & 21 & Nucleus \\
NGC 4038/4039 & Int. & 11:59:21.1 & -18:36:17 & 1510 & 21 & overlap region \\
Mrk 231 & Merger & 12:54:05.0 & 57:08:39 & 12650 & 170 & \\
Arp 220 & Merger & 15:32:46.9 & 23:40:08 & 5450 & 77 &\\
NGC 6090 & Int. & 16:10:24.0 & 52:35:11.0 & 8831 & 115 &\\
IRAS10565+2448 & Merger & 10:56:36.2 & 24:48:40 & 12923 & 165 & \\
IC 10 & Irr & 00:17:44.0 & 59:00:48 & -344 & 1 & 2 positions \\
He 2-10 & Irr & 08:34:07.2 & -26:14:06 & 873 & 9.2 & Nucleus \\
\hline
\end{tabular}
\end{table}

\newpage

\begin{table}
\caption {CI(1-0), CO(1-0) and $^{13}$CO(2-1) observed line strengths}
\begin{tabular}{|lcrrrr|}
\hline
 Source & offset & CI (1-0) & CO(1-0) & $^{13}$CO(2-1) & Ref \\
& arcsec & Kkms$^{-1}$ & Kkms$^{-1}$ & Kkms$^{-1}$ & \\
\hline
 NGC 891 & 0 & $14.4 \pm 1.9$ & $143 \pm 13$ & $7.0 \pm 1.4$ & 1 \\
         & 20 & $8.1 \pm 1.5$ & $130 \pm 13$ & $4.2 \pm 0.7$ & 1 \\
         & 40 & $8.3 \pm 1.0$ & $70 \pm 6$ & $2.1 \pm 0.4$ & 1 \\
         & 60 & $10.2 \pm 1.9$ & $62 \pm 6$ & $3.8 \pm 0.7$ & 1 \\
        & 80 & $13.2 \pm 1.9$ & $84 \pm 6$ & $2.8 \pm 0.7$ & 1 \\
        & 100 & $9.4 \pm 1.5$ & $58 \pm 6$ & $2.2 \pm 0.3$ & 1 \\
        & 120 & $9.4 \pm 1.3$ & $52 \pm 6$ & $2.8 \pm 0.7$ & 1 \\
        & 140 & $7.2 \pm 1.0$ & $46 \pm 4$ & $1.7 \pm 0.3$ & 1 \\
        & 180 & $5.3 \pm 1.0$ & $22 \pm 3$ & $0.9 \pm 0.2$ & 1 \\
        & 220 & $2.3 \pm 0.8$ & $20 \pm 3$ & $0.4 \pm 0.1$ & 1  \\
        & 260 & $0.2 \pm 0.2$ & $10 \pm 3$ & $0.3 \pm 0.3$ & 1 \\
\hline 
M~51 & 0,0 & $13 \pm 2$ & $45 \pm 5$ & & 2 \\
      & -24,-24 & $12 \pm 1.5 $ & $78 \pm 7$ & & 2 \\
      & 0,12 & $ 9 \pm 1.5$ & $47 \pm 5$ & & 2 \\
\hline 
NGC~4736 & 0,0 & $12 \pm 4$ & $32 \pm 3$ & $1.8 \pm 0.7$ & 3 \\
         & 40,0 & $2.6 \pm 0.6$ & $17 \pm 2$ & $0.8 \pm 0.15 $ & 3 \\
\hline 
NGC~3079 & 0,0 & $53 \pm 6$ & $231 \pm 10$ & $ \pm $ & 4 \\
         & 10,-50 & $4.0 \pm 1.0$ & $25 \pm 2$ & $ \pm $ & 4 \\
         & -35,130 & $1.7 \pm 0.5$ & $1.4 \pm 0.5$ & $ \pm $ & 4 \\
\hline
\end{tabular}
\end{table}

\newpage
\begin{table}
\begin{tabular}{|lcrrrr|}
\hline 
Source & offset & CI (1-0) & CO(1-0) & $^{13}$CO(2-1) & Ref \\
& & Kkms$^{-1}$ & Kkms$^{-1}$ & Kkms$^{-1}$ &  \\
\hline
 NGC~6946 & 0,0 & $53 \pm 5.7$ & $150 \pm 12$ & $\pm $ & 5 \\
          & 20,0 & $25 \pm 3.8$ & $60 \pm 6$ & $\pm $ & 5 \\
          & 40,0 & $4.5 \pm 1.3$ & $16 \pm 2.5$ & $\pm $ & 5 \\
          & 60,0 & $2.8 \pm 1.0$ & $ \pm $ & $\pm $ & 5 \\
          & 80,0 & $6.2 \pm 1.5$ & $ \pm $ & $\pm $ & 5 \\
          & 100,0 & $1.9 \pm 1.5$ & $9.6 \pm 1.2$ & $ \pm $ & 5 \\
          & 120,0 & $3.8 \pm 1.0$ & $ 7.8 \pm 1.2$ & $ \pm $ & 5 \\
          & 140,0 & $1.3 \pm 0.6$ & $12 \pm 1.8$ & $ \pm $ & 5 \\
          & 150,-20 & $3.8 \pm 1.0$ & $19 \pm 2.4$ & $\pm $ & 5 \\
          & 110,100 & $4.2 \pm 1.0$ & $ \pm $ & $ \pm $ & 5 \\
\hline
 He 2-10 & 0,0 & $3.8 \pm 0.6$ & $30 \pm 4$ & $0.4 \pm 0.1$ & 6 \\
IC~10 & 5,-20 & $5.5 \pm 1.0$ & $22 \pm 3$ & $1.0 \pm 0.15$ & 7 \\
      & -30,240 & $1.3 \pm 0.6$ & $10 \pm 1.5$ & $0.35 \pm 0.07$ & 7 \\
\hline 
NGC~4038/39 & overlap & $ 9.5 \pm 1.9$ & $102 \pm 20$ & $2.7 \pm 0.4$ & 8 \\
NGC~4038 & 0,0 & $7.6 \pm 1.9$ & $70 \pm 10$ & $2.8 \pm 0.7$ & 8 \\
Arp~220 & & $19 \pm 2.8$ & $109 \pm 10$ & $8.5 \pm 1$ & 9 \\
Mrk~231 & & $3.8 \pm 1.0 $ & $22 \pm 3$ & $ \pm $ & 10 \\
NGC~6090 & & $5.1 \pm 1.0$ & $14 \pm 3$ & $ \pm $ & 11 \\
IRAS~10565+2448 & & $9.5 \pm 2.8$ & $16 \pm 2$ & $ \pm $ & 10 \\
\hline 
Centaurus A & 0,0 & $81 \pm 7.6$ & $100 \pm 20$ & $7.5 \pm 3.0$ & 12 \\
\hline
\end{tabular}

{\small (1) Garcia-Burillo et al.  (\cite{gg}), 
(2) Garcia-Burillo et al. (\cite{gg1}), 
(3) Gerin et al.  (\cite {gcc}),
 (4) Braine et al.  (\cite {braine}),
 (5) Casoli et al. (\cite {casoli}), 
(6) Baas et al. (\cite {baas}),
(7) Becker  (\cite {becker}), 
(8) Aalto et al. (\cite {aalto}), 
(9) Radford et al. (\cite {radford}), 
(10) Solomon et al. (\cite {solomon}), 
(11) Liszt (\cite {liszt}),
 (12) Isra\"el et al. (\cite {israel})
}
\end{table}

\newpage
\begin{table}
\caption {Contribution of CI(1-0) and CO lines to the thermal balance }
\begin{tabular}{|lccc|}
\hline
 Source & CI (1-0) & CO$^1$ & FIR \\
       & 10$^{-17}$ Wm$^{-2}$ & 10$^{-17}$ Wm$^{-2}$ &
 10$^{-12}$ Wm$^{-2}$ \\
\hline 
NGC~891 - Center & 1.2 & 2.4 & 4.5 \\
NGC~891 - (100'' NE) & 0.82 & 1.3 & \\
M~51$^*$ - Center & 1.1 & 3.3 & 6.7 \\
M~51$^*$ - (-24'',-24'') & 1.0 & 2.6 & \\
NGC~4736 - Center & 1.0 & 2.0 & 3.7 \\
NGC~4736 - (40'',0) & 0.23 & 0.95 & \\
NGC~6946$^*$ - Center & 4.6 & 21 & 8.8 \\
NGC~6946 - (150'',-20'') & 0.33 & 1.1 & \\
NGC~3079 - Center & 4.6 & $ \geq 2.1$ & 2.9 \\
NGC~3079 - (-35'',130'') & 0.15 & $\geq 0.02$ & \\
\hline 
He2-10 & 0.33 & 1.1 & 1.1 \\
 IC~10$^*$ & 0.48 & 2.3 & 2.0 \\
\hline 
Centaurus A & 7.1 & 6.1 & 13.4 \\
NGC 4038/4039 & 0.82 & 3.9 & 2.2 \\
NGC 6090 & 0.44 & 1.2 & 0.33 \\
Arp 220 & 1.6 & 2.7 & 4.7 \\
Mrk 231 & 0.33 & 0.56 & 1.4 \\
\hline
\end{tabular}

{\small $^1$ total CO flux from observed lines : (1-0), (2-1), (3-2) and
(4-3) when available. \\
Galaxies with CO(4-3) data are indicated with an
asterisk.}
\end{table}

\begin{thebibliography}{}
\bibitem [1995] {aalto} Aalto A., Booth R.S., Black J.H., Johansson
L.E.B., 1995, A\&A 300, 369.
\bibitem [1992]{aa} Abgrall H., Le Bourlot J., Pineau des For\^ets G.,
Roueff E., Flower D.R., Heck L., 1992, A\&A 253, 525. 
\bibitem [1994] {baas} Baas F., Israel F.P., Koornneef J., 1994, 
A\&A 284, 403
\bibitem [1998]{bt} Bakes E.L.O., Tielens, A.G.G.M., 1998, ApJ 499, 258.
\bibitem [1992] {becker} Becker R., PhD Thesis.
\bibitem [1994]{bennett} Bennett C.L., Fixsen D.J., Hinshaw G. et al., ApJ
434, 587. 
\bibitem [1997] {braine} Braine J.,Gu\'elin M., Dumke M., Brouillet N.,
 Herpin F., Wielebinski R., 1997, A\&A 326, 963. 
\bibitem [1998]{1998} Buat V., Burgarella D. 1998, A\&A 334, 772. 
\bibitem [1992]{bkp} B\"uttgenbach T.H., Keene J., Phillips T.G., 
Walker C.K., 1992, ApJ 397, L15. 
\bibitem [1990] {casoli} Casoli F., Clausset F., Combes F., Viallefond F.,
Boulanger F., 1990, A\&A 233, 357. 
\bibitem [1992] {casoli2} Casoli F., Dupraz C., Combes F., 1992, A\&A
264, 55
\bibitem [1999]{colbert} Colbert J.W., Malkan M.A., Clegg P.E. et al.,
1999, ApJ 511, 721.
\bibitem [1989] {frerking} Frerking M.A., Keene J., Blake G.A., Phillips
T.G., 1989, ApJ 344, 311.
\bibitem [1992] {gg} Garcia-Burillo S., Gu\'elin M., Cernicharo J.,
 Dahlem M., 1992, A\&A 266, 21.
\bibitem [1993] {gg1} Garcia-Burillo S., Gu\'elin M., Cernicharo J., 1993, 
A\&A 274, 123.
\bibitem [1991] {gcc} Gerin M., Casoli F., Combes F., 1991, A\&A 251, 32. 
\bibitem [1997] {gp2} Gerin M., Phillips T.G., 1997, 
{\it The Far Infrared and Submillimetre Universe}, ESA SP-401, p 105. 
\bibitem [1998] {gp1} Gerin M., Phillips T.G., 1998, ApJ 509, L17. 
\bibitem [1993] {guelin} Gu\'elin M., Zylka R., Mezger P.G., Haslam C.G.T.,
Kreysa E., Lemke R., Sievers A.W., A\&A 279, L37.
\bibitem [1993]{gusten} G\"usten R., Searbyn E., Kasemann C. et al., 1993,
 ApJ 402, 537.
\bibitem [1995]{harrison} Harrison A., Puxley P., Russel A., Brand P., 1995,
MNRAS 277, 413.
\bibitem [1991] {israel} Israel F.P., van Dishoeck E.F.,Baas F., De Graauw
T., Phillips T.G., 1991, A\&A 245, L13. 
\bibitem [1998]{itb} Israel F.P., Tilanus R.P.J., Baas F., 1998, A\&A 339, 398.
\bibitem [1995]{iwb} Israel F.P., White G.J., Baas F., 1995, A\&A 302, 343. 
\bibitem [1979]{}Jenkins E.B., Shaya E.J., 1979, ApJ 231, 55.
\bibitem [1999]{}Kaufman M.L., Wolfire M.G., Hollenbach D.J., Luhman M.L.,
1999 ApJ submitted.
\bibitem [1996] {keene} Keene J., Lis D.C., Phillips T.G., Schilke P., 1996,
IAU 178, p 129. 
\bibitem [1993]{}Le Bourlot J., Pineau des For\^{e}ts G., Roueff E., 
Flower D.R., 1993,A\&A 267, 233.
\bibitem [1994] {} Lequeux J., Le Bourlot J., Pineau des For\^ets G.,
Roueff E., Boulanger F., Rubio M., 1994, A\&A 292, 371.
\bibitem [1992]{liszt} Liszt H.S., 1992, ApJ 386, 139. 
\bibitem [1998]{} Luhman M.L., Satyapal S., Fisher J.,
Wolfire M.G., Cox P., Lord S.D., Smith H.A., Stacey G.J.,
Unger S.J., 1998, ApJ 504, L11.
\bibitem [1983]{mmp} Mathis J.S., Mezger P.G., Panagia, N.1983, A\&A 128, 212.
\bibitem [1995] {mw} Minchin N.R., White G.J., 1995, A\&A 302, L25.
\bibitem [1998] {} Pak S., Jaffe D.T., van Dishoeck E.F., Johansson
L.E.B., Booth R., 1998, ApJ 498, 735. 
\bibitem [1998]{pw} Petitpas G.R., Wilson C.D., 1998, ApJ 503, 219. 
\bibitem [1981]{}Phillips T.G., Huggins P.J, 1981, ApJ 251, 533. 
\bibitem [1999] {} Plume R., Jaffe D.T., Tatematsu K., Evans N.J.,
Keene J., 1999, ApJ 512, 768.
\bibitem [1991] {radford} Radford S.J.E.,Downes D., Solomon P.M.,1991, ApJ
368, L15 
\bibitem [1998] {sauty} Sauty S., Gerin M., Casoli F., 1998, A\&A 339, 19. 
\bibitem [1993]{schilke} Schilke P., Carlstrom J.E., Keene J., Phillips T.G.,
1993, ApJ 417, L67.
\bibitem [1996] {1996}Smith B.J., Harvey P.M., 1996, ApJ 468, 139.
\bibitem [1995] {}Sodroski T.J., et al., 1995, ApJ 452, 262.
\bibitem [1997]{solomon} Solomon P.M., Downes D., Radford S.J.E., Barrett
J.W., 1997, ApJ 478, 144.
\bibitem [1991] {1991}Stacey G.J., Geis N., Genzel R., 1991, ApJ 373, 423.
\bibitem [1997]{starck} Starck A.A., Bolatto A.D., Chamberlin R.A., Lane
A.P., Bania T.M., Jackson J.M., Lo K., 1997, ApJ 480, L59.
\bibitem [1996]{stark-r} Stark R., Wesselius P.R., van Dishoeck E.F.,
Laureijs R.J., 1996, A\&A 311, 282.
\bibitem [1997]{stutzki} Stutzki J., Graf U.U. et al., 1997, ApJ 477, L33.
\bibitem [1995] {tauber} Tauber J., A., Lis D.C., Keene J., Schilke P., 
B\"uttgenbach T.H., 1995, A\&A 297, 567. 
\bibitem [1985]{} Tielens A.G.G.M., Hollenbach D., 1985,
ApJ 291, 722.
\bibitem [1995]{ws} White G.J., Sandell G., 1995, A\&A 299, 179.
\bibitem [1997]{wilson} Wilson C.D., 1997, ApJ 487, L49.
\end{thebibliography}
\end{document}